\documentclass{article}
\usepackage{spconf,amsmath,graphicx,hyperref}
\usepackage{comment}
\usepackage{booktabs}
\usepackage{booktabs}
\usepackage{siunitx}
\usepackage{stfloats}

\title{Acoustic and Facial Markers of Perceived Conversational Success in Spontaneous Speech}
\name{Thanushi Withanage\thanks{This work is supported by Grand Challenge grant from University of Maryland}$^{\dagger}$ \qquad Elizabeth Redcay$^{\star}$ \qquad Carol Espy-Wilson$^{\dagger}$}
  
  \address{$^{\dagger}$Department of Electrical and Computer Engineering, University of Maryland College Park, MD, USA  \\
      $^{\star}$Department of Psychology, University of Maryland College Park, MD, USA }

\begin{document}
%
\maketitle

\begin{abstract}

Individuals often align their speaking patterns with their interlocutors, a phenomenon linked to engagement and rapport. While well documented in task-oriented dialogues, less is known about entrainment in naturalistic, non-task and virtual settings. In this study, we analyze a large corpus of spontaneous dyadic Zoom conversations to examine how conversational dynamics relate to perceived interaction quality. We extract multimodal features encompassing turn-taking, pauses, facial movements, and acoustic measures such as pitch and intensity. Perceived conversational success was quantified via factor analysis of post-conversation ratings. Results demonstrate that entrainment reliably detected in spontaneous speech and correlates with higher perceived success. These findings identify key interactional markers of conversational quality and highlight opportunities for targeted interventions to foster more effective and engaging communication. 
\end{abstract}

\vspace{-0.2cm}
\vspace{-0.4cm}

\section{Introduction}
\label{sec:intro}

\vspace{-0.2cm}
Social interaction is vital for communication, bonding, and well-being. Partners influence each other across speech, language, and visual cues \cite{multimodal,multi}. In spontaneous \textit{get-to-know-you} conversations, turn dynamics may reveal interaction quality: shorter turns may reflect alignment or disengagement, while longer turns can indicate comfort and topic engagement. Notably, longer gaps foster connection among friends but have shown disconnection among strangers \cite{gaps}. 

In conversation, interlocutors adapt by mirroring emotions, prosody, and body movements, a process termed alignment, entrainment, or mimicry \cite{ent1}. High entrainment, measured as time-locked behavioral alignment, enhances interaction quality \cite{ent7,candor}. Both verbal and non-verbal cues are critical for successful interaction \cite{multi}, which further depends on shared attentiveness, coordination, and mood. Research shows individuals can learn to entrain through targeted training, for example in children with speech difficulties \cite{train}. Identifying alignment features that predict conversational success is therefore vital for developing effective training methods.

Entrainment of acoustic–prosodic features such as pitch, intensity, speaking rate, and loudness has been shown to predict conversation quality in task-oriented settings \cite{ent7,ent9,ent6}. Non-verbal features including body movements, eye gaze, facial expressions, and head movements also predict interaction quality \cite{sm,au}. However, a major gap remains in understanding unstructured, naturalistic conversations between adult peers in extended Zoom interactions. Prior research has largely focused on task-oriented contexts, such as computer games \cite{game}, collaborative problem solving in education \cite{ent7}, and user–agent interactions \cite{agent}, primarily through acoustic–prosodic features. Other studies have examined face-to-face dynamics in spontaneous code-switching \cite{code} or short get-to-know-you interactions with autistic children \cite{aut1}, but the focus here is on adult peers conversing in the same language in long virtual settings. 

Most prior spontaneous conversation studies were in-person, yet many contemporary peer interactions including job interviews now occur remotely. Understanding how entrainment during remote communication aligns with conversational success is therefore critical. Our study investigates whether multimodal entrainment predicts self-reported enjoyment and alignment, with the long-term goal of identifying markers of conversation quality in both neurotypical and mixed-neurotype interactions. The main contributions of this work are; we show that features; facial action units, pitch, intensity, turn count, pause duration are associates with perceived conversational success. We analyzed a large corpus of naturalistic conversations held via Zoom and show entrainment occurs and it correlates with conversation success. In this work, we are focusing on conversational alignment. 

First, we initialize our study by observing the general conversation dynamics like turn and pause trend of conversations. Next utilizing features such as facial expressions, pitch and intensity we fine-tune our analysis into temporal window-based and turn-based analysis to investigate if there is speaker entrainment occurring across time. Finally, we aim to see whether there is an association of the aforementioned features with the perceived conversation success.
\vspace{-0.55cm}
\section{Dataset}
\label{sec:format}
\vspace{-0.3cm}
For this study we analyze the CANDOR Corpus (Conversation: A Naturalistic Dataset of Online Recordings) \cite{candor}, collected by BetterUp Labs in collaboration with researchers at the University of Pennsylvania (2023). CANDOR comprises more than 1500 spontaneous, dyadic, ~30-minute video and audio recorded conversations conducted over Zoom between previously unacquainted adults aged 19–66 who represent a broad range of gender, educational, ethnic, and generational identities. Interactions were unstructured and non-task-oriented. After each conversation, both participants independently completed a post-conversation survey rating multiple aspects of perceived interaction quality; these ratings were used to construct a composite success score (section \ref{ssec:pcs}). Separate audio channels were recorded for each participant, obviating the need for speaker diarization. For all analyses, we restricted the sample to sessions flagged (by the Dataset) as free of background noise or interruptions.

\vspace{-0.5cm}

\section{Method}
\label{sec:pagestyle}
\vspace{-0.3cm}
\subsection{Perceived conversation success}
\label{ssec:pcs}
\vspace{-0.2cm}
Each participant completed both pre and post conversation survey questionnaires designed to assess the quality of the interaction. The full instrument contained 229 items, including demographic and speaker-related details. For the purposes of this study, we focused on a subset of 21 constructs that capture dimensions of affect, enjoyability, friendship, and common ground. To identify relevant dimensions of perceived conversational success (PCS), we conducted a principal component analysis (PCA) on these constructs. An initial exploratory PCA revealed two latent dimensions with loadings exceeding 0.4. Accordingly, we performed a subsequent PCA constrained to two dimensions ($PCA_1$, $PCA_2$), retaining constructs with loadings above 0.4. Although both components were interpretable, only $PCA_1$ was adopted as the basis for the PCS measure. This choice was motivated by its stronger discriminative utility: when analyzed separately, constructs grouped under $PCA_1$ exhibited significantly greater separation on PCS in terms of feature-level entrainment, whereas $PCA_2$ did not yield comparable distinctions.

Responses to those 11 constructs in $PCA_1$; \textit{Affect, Overall affect, Affect at beginning, Affect at middle, Affect at end, Best affect, How much enjoyable, I like you, You like me, Conversationalist, My friends like you} were originally recorded on heterogeneous scales (1–7, 1–9, or 1–100) and normalized to a common range prior to analysis. Ratings were z-score normalized within each construct and averaged across constructs to yield an individual PCS score bounded between 0 and 1. To reduce label ambiguity and enable a high-contrast assessment of entrainment-related discriminative validity, we focused on conversations at the extreme ends of the PCS distribution, retaining only those more than one standard deviation from the median. Given the overall high enjoyment levels, this corresponded to PCS $\leq 0.6 $ for Low-Successful Conversations (LSCs) and PCS $\geq 0.9 $ for High-Successful Conversations (HSCs), resulting in 35 LSCs and 91 HSCs.
\vspace{-0.3cm}

\subsection{Turn taking analysis}
\label{ssec:subhead}
  \vspace{-0.2cm}
While turn-taking is a well-established conversational norm \cite{gaps}, the fine-grained structure of turn exchanges is complex, and varied patterns can nonetheless produce highly successful interactions. To test these hypotheses, we derived turn-level measures from the dataset’s $Backbiter$ transcripts, which exclude backchannel utterances from turn units and thus provide a clearer operational definition of conversational turns. While no single definition of a $turn$ is universally accepted, we follow the dataset’s convention (turn is defined as a contiguous stretch of speech by one speaker bounded by a change of floor) for computational consistency \cite{candor}. 

For each conversation (including turns from both speakers), we computed summary statistics of turn duration; minimum, maximum, mean, total (sum across all turns), and overall turn count using the annotated start and end times from the transcripts. We then examined their association with PCS.

In parallel, we quantified inter-turn silence, defined as the pause (silences exceeding 0.6 s considered significant \cite{candor}) between one speaker’s offset and the other speaker’s onset. Using the same timing annotations, we calculated the minimum, maximum, mean, and total pause duration per conversation and evaluated their relationship to PCS.
  \vspace{-0.3cm}
\subsection{Acoustic features analysis}
\label{ssec:turn}
\begin{figure*}[t]
  \centering
  \begin{minipage}{0.42\linewidth}
    \centering
    \includegraphics[width=\linewidth]{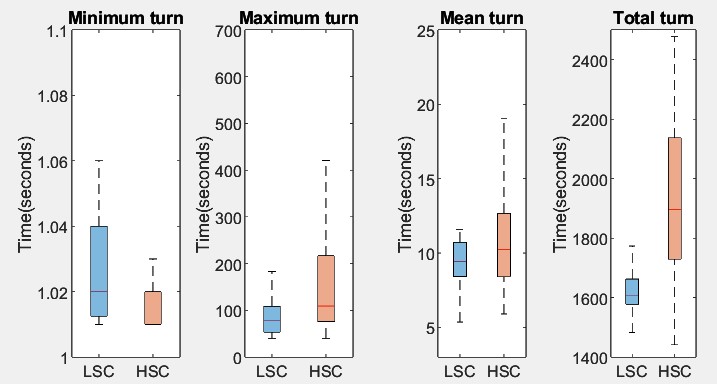}\\
    (a) Turn duration statistics
  \end{minipage}
  \hfill
  \begin{minipage}{0.42\linewidth}
    \centering
    \includegraphics[width=\linewidth]{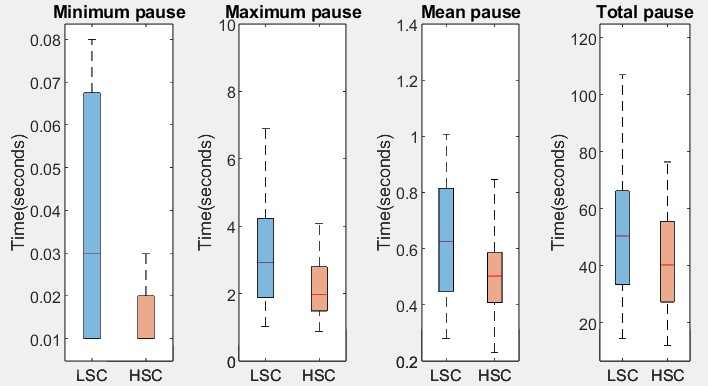}\\
    (b) Pause duration statistics
  \end{minipage}
  \hfill
  \begin{minipage}{0.1125\linewidth}
    \centering
    \includegraphics[width=\linewidth]{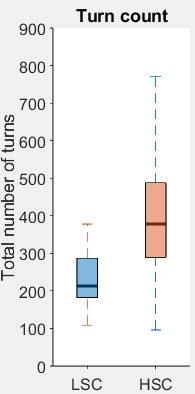}\\
    (c) Turn count
  \end{minipage}
    \caption{$f$ vs. PCS. (a) Turn duration (min, max, mean, total). 
             (b) Pause duration (min, max, mean, total). 
             (c) Turn count.}
  \label{fig:turnpause}
\end{figure*}

Each speaker’s audio was recorded on an independent channel. All audio recordings were first converted from dual-channel to mono and downsampled from 44.1 kHz to 16 kHz to standardize the signal representation. Features were segmented into speaker turns based on transcript-provided timing boundaries. 

To analyze acoustic dynamics, we extracted the pitch (\(F0\)) from each speaker’s turn-level audio using the Pitch estimating Neural Network (PENN) \cite{pen}, which has demonstrated robust performance, including detection of \(F0\) during creaky-voice regions. Pitch trajectories were subsequently normalized to reduce gender-related variability. In addition, speech intensity was computed for each turn using $Praat$. For the statistical analysis minimum, maximum and mean of each turn's acoustic feature were computed and their entrainment along the conversation duration is tracked.
  \vspace{-0.3cm}

\subsubsection{Turn level proximity entrainment in acoustic features}
\label{sssec:subsubhead}

To quantify acoustic entrainment, we calculate turn-level proximity\cite{game} which is the phenomenon where adjacent turns should lie in close proximity compared to non-adjacent turns. First, for each conversation, we indexed the turns in temporal order \(i=1,...,N\). For a given acoustic feature statistic \(f\) (we use turn-level summary statistics: minimum, maximum, and mean of \(F0\) and intensity), we calculated the feature value for the current speaker on turn \(i\) denoted by \(fc_i\), and the corresponding feature value for the partner on the next turn \(fp_{(i+1)}\).
Adjacent-turn distance is defined as the absolute difference between those two as in equation \ref{equation:eq1}
  \vspace{-0.3cm}
\begin{align}
  fd_a(i) &= \lvert fc_i - fp_{(i+1)}\rvert 
  \label{equation:eq1}
\end{align}
 
To obtain a non-adjacent baseline,we randomly choose a partner turns' \(f\) that is non-adjacent to the turn \(i\) (\(fp_{j\neq i}\)) and took the absolute difference of that with \(fc_i\). This process is repeated a total 10 times and average the of those 10 differences is computed as shown in equation \ref{equation:eq2}
  \vspace{-0.1cm}
\begin{align}
fd_{na}(i) &= \frac{\Sigma^{10}_{j}\lvert fc_i - fp_{j\neq i}\rvert }{10}
  \label{equation:eq2}
\end{align}
which we refer to as the non-adjacent distance. If entrainment is present, adjacent differences \(fd_a(i)\) should be smaller than their non-adjacent counterparts \(fd_{na}(i)\). 
We compute \(fd_a(i)\) and \(fd_{na}(i)\) for every turn in every conversation and for each feature statistic \(f\). The resulting distance distributions were compared using the Mann–Whitney U test to evaluate whether adjacent-turn distances were systematically smaller than non-adjacent baselines. Shapiro–Wilk tests revealed significant deviations from normality across features, particularly in HSCs. For example, minimum pitch showed strong non-normality in both groups $(LSC: p=5.15e-8, HSC: 4.16e-17)$. Given this pattern, the Mann–Whitney U test was applied for all analyses.
\vspace{-0.3cm}

\subsection{Facial expression analysis}
\label{ssec:subhead}
To study entrainment of facial expressions we extract Facial Action Units (FAUs) using the OpenFace open-source facial behavior analysis toolkit \cite{opf}, which processes video recordings of each speaker and outputs a wide range of facial movement indicators. For this study, we used the 17 FAUs (mentioned in Table~\ref{tab:fau}) extracted from the default OpenFace settings. 

  \vspace{-0.3cm}
\subsubsection{Synchrony in facial action units}
\label{sssec:subsubhead}

While proximity reflects the extent to which a given feature is similar in magnitude across interlocutors, synchrony captures the temporal alignment of feature trajectories, even when their absolute values may differ.

In this work, we investigate whether the synchrony of FAUs aligns with PCS. Specifically, we computed Pearson correlations between the same FAU across the two speakers within each conversation, using non overlapping 5 second windows.  This continuous, fixed-window approach was chosen in place of turn-based segmentation, as meaningful emotional expressions can also occur during pauses, which would otherwise be excluded in turn-based analysis. \cite{ent6}

When participants mirror each other’s facial expressions, the correlation between their facial action units increases. To capture this phenomenon, raw correlation values were first converted to Fisher z-transformed values ($z_f$). We then computed the mean of $z_f$ across each conversation. This measure was calculated independently for each FAU and subsequently examined in relation to PCS ratings.
  \vspace{-0.3cm}
\section{Results}
\label{sec:pagestyle}
  \vspace{-0.3cm}
\subsection{Turn and pauses vs. PCS}
\label{ssec:subhead}
To assess group differences, we conducted Mann–Whitney U tests for each \(f\) with the null hypothesis (H0) that distributions are similar and alternative hypothesis (H1) that the ditributions differ. The corresponding $U, z, p, q$ (Benjamini–Hochberg False Discovery Rate corrected value) reported in Table~\ref{tab:turn}. Figure~\ref{fig:turnpause}(\(a\)) shows that in HSCs, maximum, mean, and total turn durations were significantly greater than in LSCs. This suggests that successful conversations are characterized by longer turns, with fewer interruptions or pauses between turns. 

To further support this interpretation, we analyzed pause durations. The results in Figure~\ref{fig:turnpause}(\(b\)) indicated that HSCs had shorter and less frequent pauses than LSCs. Mann–Whitney U tests confirmed this pattern (see Table~\ref{tab:turn}). 

The total number of turns (across both speakers shown in Figure~\ref{fig:turnpause}(\(c\))) also differed markedly between HSC and LSC conversations. A Mann–Whitney U test revealed a strong class effect $(U= 545, z= -5.71, p= 1.18e-8)$, showing that turn counts were significantly lower in LSCs. The large negative
z-score underscores the robustness of this finding, suggesting that high-success conversations are marked by greater turn-taking activity. This supports the view that higher engagement and reciprocity, reflected in frequent turn exchanges, are key indicators of conversational success.

  \vspace{-0.10cm}
\begin{table*}[!t]
  \caption{Turn and pause statistics significance ($p<0.05$ is statistically significant $^*$)}
  \label{tab:turn}
  \centering
  \begin{tabular}{lllll}
    \toprule
    \textbf{\(f\)}   & \textbf{minimum \(f\) $(U, z, p, q)$}     & \textbf{maximum \(f\) $(U, z, p, q)$}  & \textbf{mean \(f\) $(U, z, p, q)$} & \textbf{total \(f\) $(U, z, p, q)$}\\
    \midrule
    Turn duration     &2129, 2.922, 1.54e-03,       &669, -3.037, 2.41e-03,   &522, -1.401, 1.62e-01,  &151, -5.365,8.32e-08,\\
        &\textbf{3.094e-03}$^*$  &\textbf{3.220e-03}$^*$    &1.626e-01  &\textbf{3.330e-07}$^*$  \\
    Pause duration      &2272.5, 3.704, 1.52e-05,        &1224, 2.845, 4.49e-03,   &2080, 2.655, 7.98e-03,   &1552, 1.802, 7.21e-02,\\
        &\textbf{6.0840e-05}$^*$   &\textbf{8.9880e-03}$^*$   &\textbf{1.0652e-02}$^*$   &7.2080e-02 \\
    F0      &1073, -2.826, 4.70e-03,       &1452, -0.762, 4.46e-01,   &1213, -2.064, 3.90E-02,  &-\\
        &\textbf{1.41e-02}$^*$  &4.46e-01   &5.85e-02   \\
  
    Intensity       &1008, -2.286, 2.22e-02,        &920, -2.819, 4.81e-03,   &700, -4.152, 3.30e-05,  &- \\
        &\textbf{2.22e-02}$^*$   &\textbf{7.21e-03}$^*$   &\textbf{9.89e-05}$^*$   \\
    \bottomrule
  \end{tabular}
\end{table*}

\subsection{Synchrony in facial expressions vs. PCS}
\label{ssec:subhead}

To examine facial synchrony, we computed windowed correlations for each FAU, comparing the same FAU across both speakers within each conversation. Because the Fisher z-transformed correlation values $z_f$ are normally distributed, we employed Welch’s t-tests to compare synchrony values between LSCs and HSCs. In addition, we calculated temporal statistics, including the mean and standard deviation of $z_f$ for each FAU (see Table~\ref{tab:fau}). FDR correction of p-values yielded $q=[ 0.33,0.42,0.49 ]$ for the first three prominent FAUs. 

Across the 17 FAUs analyzed, synchrony in those associated with smiling expressions \cite{Cohn}; particularly the first five FAUs linked to smiling was consistently higher in HSCs than in LSCs. These findings suggest that synchrony in positive affective expressions, such as smiling, constitutes a robust nonverbal marker of perceived conversational success. In contrast, synchrony in FAUs associated with negative affect contributed little to PCS and, when present, tended to be greater in LSCs as observed from \(t\) statistic in Table~\ref{tab:fau}).
  \vspace{-0.3cm}

\subsection{Proximity in F0 and intensity}
\label{ssec:subhead}

For each speaker turn in a conversation, we extracted two distance measures \(fd_{a}(i)\) and \(fd_{na}(i)\) for each acoustic \(f\) (pitch and intensity) across three statistics (minimum, maximum, and mean). This procedure yielded paired sets of distances (\(fd_{a}(i)\), \(fd_{na}(i)\)).  To evaluate whether adjacent-turn distances were systematically smaller than non-adjacent distances, we applied a paired t-test within each conversation. The null hypothesis (H0) was that the means of \(fd_{a}(i)\) and \(fd_{na}(i)\) were equal, while the alternative hypothesis (H1) stated that these means differed. From each test, we also computed Cliff’s delta ($\delta$) to estimate the effect size, quantifying the magnitude and direction of proximity entrainment.

These per-conversation effect size values were then compared between HSC and LSC using Mann–Whitney U tests (summarized in Table~\ref{tab:turn}). For example, Cliff’s delta values for pitch were $\delta = [-0.326,\,-0.088,\,-0.238]$ for minimum, maximum, and mean statistics, respectively, while for intensity they were $\delta = [-0.273,\,-0.336,\,-0.495]$. These negative values indicate stronger entrainment in HSCs, consistent with significantly smaller adjacent-turn distances relative to the non-adjacent baseline. Mann–Whitney results confirmed this pattern: minimum pitch, and minimum, maximum, and mean intensity showed significant differences between HSC and LSC conversations, highlighting proximity entrainment as a reliable acoustic marker of conversational success.

\begin{table}[t]
\centering
\caption{Welch's $t$-test results for Fisher-transformed FAU correlations. Values show $t$, $p$,  and mean correlation ($\mu$) with standard deviations ($\sigma$) for LSC and HSC. Bolded $p<0.1$ $^*$}
\scriptsize
\begin{tabular}{lrlcc}
\toprule
\textbf{FAU ID:Description} & \textbf{$t$} & \textbf{$p$} & \textbf{($\mu_{L}$, $\sigma_L$)} & \textbf{($\mu_{H}$, $\sigma_H$)} \\
\midrule
10: Upper Lip Raiser & -2.43 & \textbf{1.96e-02}$^{*}$ & (0.35, 0.17) & (0.42, 0.12) \\
14: Dimpler  & -2.01 & \textbf{5.02e-02}$^{*}$ & (0.36, 0.14) & (0.41, 0.11) \\
07: Lid Tightener  & -1.74 & \textbf{8.80e-02}$^{*}$ & (0.28, 0.17) & (0.33, 0.16) \\
12: Lip Corner Puller & -1.51 & 1.36e-01                 & (0.37, 0.20) & (0.42, 0.20) \\
06: Cheek Raiser  & -1.28 & 2.08e-01                 & (0.35, 0.20) & (0.40, 0.19) \\
17: Chin Raiser &  1.19 & 2.41e-01                 & (0.44, 0.11) & (0.42, 0.11) \\
05: Upper Lid Raiser  &  0.95 & 3.47e-01                 & (0.31, 0.12) & (0.29, 0.13) \\
02: Outer Brow Raiser  &  0.93 & 3.57e-01                 & (0.32, 0.10) & (0.29, 0.13) \\
20: Lip Stretcher & -0.49 & 6.27e-01                 & (0.38, 0.10) & (0.37, 0.11) \\
04: Brow Lowerer  &  0.37 & 7.14e-01                 & (0.37, 0.16) & (0.34, 0.18) \\
15: Lip Corner Depressor &  0.30 & 7.63e-01                 & (0.39, 0.11) & (0.38, 0.11) \\
09: Nose Wrinkler  & -0.23 & 8.16e-01                 & (0.35, 0.12) & (0.35, 0.10) \\
45: Blink  &  0.13 & 8.94e-01                 & (0.39, 0.09) & (0.39, 0.10) \\
23: Lip Tightener &  0.12 & 9.03e-01                 & (0.39, 0.10) & (0.37, 0.11) \\
26: Jaw Drop &  0.08 & 9.35e-01                 & (0.40, 0.11) & (0.40, 0.09) \\
25: Lips part &  0.07 & 9.47e-01                 & (0.41, 0.11) & (0.40, 0.10) \\
01: Inner Brow Raiser  &  0.04 & 9.69e-01                 & (0.41, 0.11) & (0.41, 0.11) \\
\bottomrule
\end{tabular}
\label{tab:fau}
\end{table}

  \vspace{-0.15cm}

\section{Conclusion and future work}
\label{sec:pagestyle}

This study demonstrates that PCS is positively associated with multiple conversational dynamics, including turn and pause duration, facial movement synchrony, and pitch and speech intensity proximity during spontaneous online interactions. HSCs are characterized by a greater number of turns with longer durations and shorter pauses, whereas LSCs exhibit fewer, shorter turns and longer pauses. HSCs also show a higher prevalence of positive FAU correlations, indicating stronger interpersonal synchrony. These findings align with prior work and highlight the importance of situating our results within a broader interactional context. In future work we will examine these dynamics in mixed neurotype interactions, including conversations between autistic and neurotypical adults, using continuous PCS modeling.

\vfill\pagebreak

\bibliographystyle{IEEEbib}
\bibliography{Template}

\end{document}